\documentclass[aip,rsi,reprint]{revtex4-1}

\usepackage{subfigure,natbib}
\usepackage{graphicx,amsmath,amsfonts,amssymb,footnote,placeins,textcomp,epstopdf}

\newcommand\PicoMeter{\nobreak\mbox{$\;$pm}}

\newcommand\NanoMeter{\nobreak\mbox{$\;$nm}}
\newcommand\MicroMeter{\nobreak\mbox{$\;$\textmu m}}
\newcommand\MilliMeter{\nobreak\mbox{$\;$mm}}

\newcommand{\FemtoMeterPerRtHz}[0]{\nobreak\mbox{$\;$fm$\;$Hz$^{-1/2}$}}
\newcommand\Second{\nobreak\mbox{$\;$s}}

\newcommand\MicroSecond{\nobreak\mbox{$\;$\textmu s}}
\newcommand\MicroHertz{\nobreak\mbox{$\;$\textmu Hz}}
\newcommand\MilliHertz{\nobreak\mbox{$\;$mHz}}
\newcommand\Hertz{\nobreak\mbox{$\;$Hz}}
\newcommand\KiloHertz{\nobreak\mbox{$\;$kHz}}
\newcommand\MegaHertz{\nobreak\mbox{$\;$MHz}}
\newcommand\GigaHertz{\nobreak\mbox{$\;$GHz}}
\newcommand\DeciBell{\nobreak\mbox{$\;$dB}}

\newcommand\dBm{\nobreak\mbox{$\;$dBm}}

\newcommand\Volt{\nobreak\mbox{$\;$V}}
\newcommand{\NmPerK}[0]{\nobreak\mbox{$\;$nm$\;$K$^{-1}$}}
\newcommand{\PerK}[0]{\nobreak\mbox{$\;$K$^{-1}$}}

\begin{document}
\title{Construction and Characterization of a Frequency-Controlled, Picometer-Resolution, Displacement Encoder-Actuator} 

\author{John P. Koulakis}
\affiliation{$^{1)}$Department of Physics and Astronomy, University of California Los Angeles, Los Angeles, California 90095, USA}
\author{Michael Stein}
\noaffiliation
\author{K\'aroly Holczer}
\email[]{holczer@physics.ucla.edu}
\affiliation{$^{1)}$Department of Physics and Astronomy, University of California Los Angeles, Los Angeles, California 90095, USA}

\date{\today}

\begin{abstract}
We have constructed an actuator/encoder whose generated displacement is controlled through the resonance frequency of a microwave cavity. A compact, 10-\textmu m-range, digitally-controlled actuator executing frequency-coded displacement with picometer resolution is described.  We consider this approach particularly suitable for metrologic-precision scanning probe microscopy.
\end{abstract}

\maketitle
The easy mechanical tunability of microwave cavities has been exploited in all imaginable applications from broadband, mechanically-tuned microwave oscillators and filters to gravitational wave detection\cite{MannBlair} thanks to the ever growing variety of resonator structures and frequency locking schemes.  High sensitivity transducers based on microwave cavities\cite{BlairIvanovPeng,*CuthbertsonTobarIvanovBlair} have been reported as well.  In this paper we describe a displacement actuator system - a microwave cavity and controller - designed specifically for Scanning Probe Microscopy (SPM) applications.

Positioning in scanning probe microscopy represents a particular challenge due to the enormous dynamic range (${>10^7}$) of the piezo-electric actuators used.  While scan ranges of \mbox{3-100\MicroMeter} are common, sub-angstrom (${<10\PicoMeter}$) resolution is achieved reproducibly at low temperature\cite{PhysRevLett.97.177001, *ISI:000251456900024} and somewhat randomly in less controlled conditions.  A search for a reasonable way to build a metrologic instrument capable of making measurements with precision comparable to that of crystallographic data, is the motivation of this work.  The minimum requirement to achieve that, is an active-feedback-controlled distance encoder with ${\sim 10}$\MicroMeter\ working range, a ${>1}$\KiloHertz\ bandwidth response combined with picometer-scale accuracy, linearity, and long-term stability.  We demonstrate a device and control system that meets all these requirements in a smaller, simpler configuration than other methods\cite{ISI:000306831400006} under development.  In addition, it is self-calibrating; relying on the speed of light to set the distance scale. 

The resonant frequencies, $f_n$, of the ${\text{TEM}_{00n}}$ modes of a coaxial resonator are related to its length, $L$, through $f{_n=nc/2L}$.  The frequency's independence of all other dimensions makes these modes the most suitable for our purpose, in spite of the fact that they are relatively low quality factor (Q) modes.  To keep it compact, we choose to build a 15\MilliMeter\ length cavity working at its lowest frequency (n=1) mode, at 10\GigaHertz.  The practical realization required building a variable-length cavity, with a coupling that is minimally perturbed by the length change, as well as a sub-millihertz-resolution microwave control system with $10^{-12}$ stability.  Choosing the cavity frequency and allowing closed-loop control to match the resonator length, forms our frequency-controlled, displacement encoder-actuator.

\begin{figure}
		\includegraphics[]{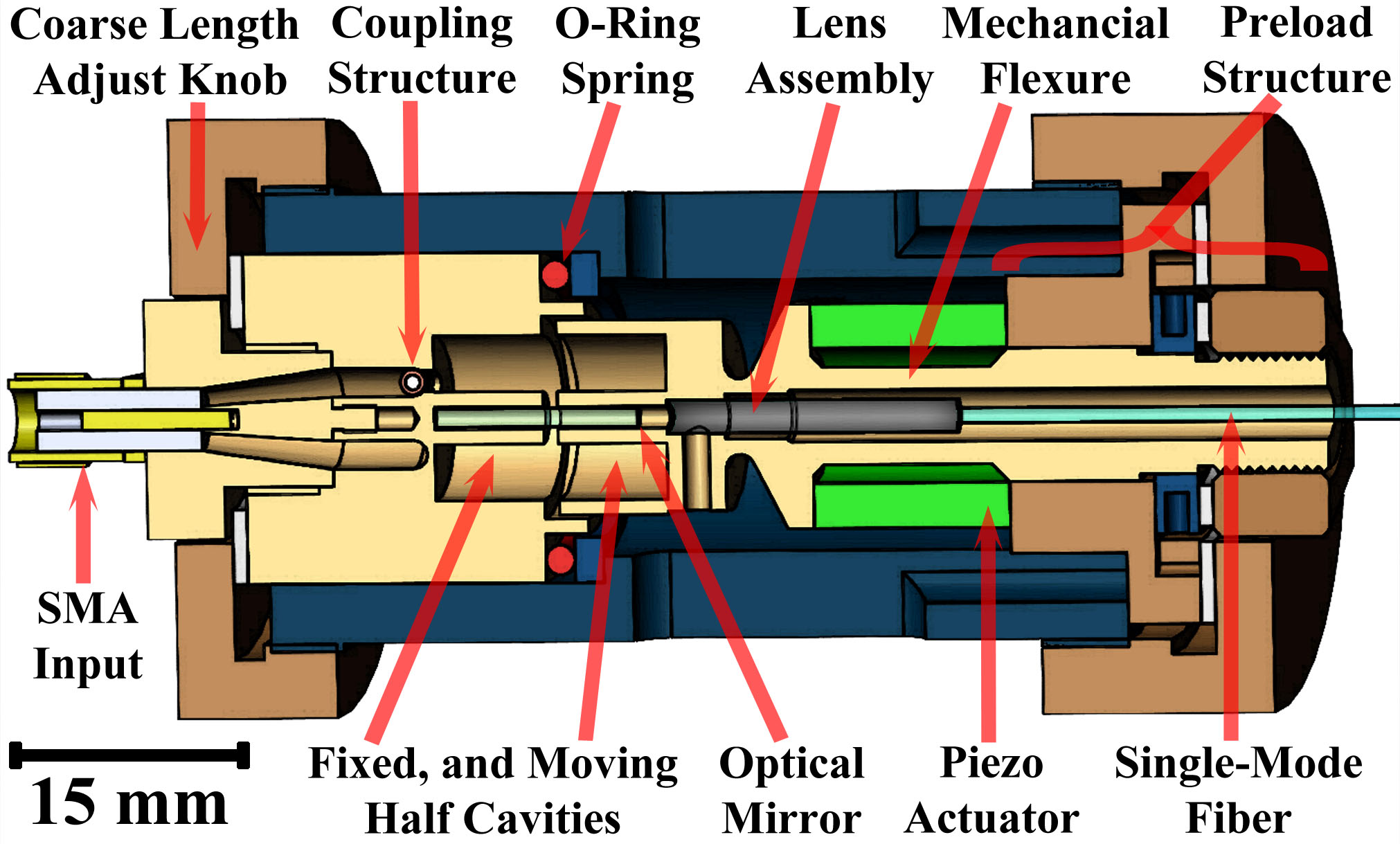}
		\caption{\label{LabeledAssembly}(Color online) Cross section of the displacement encoder assembly, which incorporates the coupling structure, the resonant cavity, a mechanical flexure, and an optical interferometer along the axis to independently measure the relative displacement of the cavity halves.  The moving half-cavity is machined into the end of an aluminum tube stretched by a concentric stacked-piezo tube.}
\end{figure}

\begin{figure*}[bt]
\includegraphics[width=\linewidth]{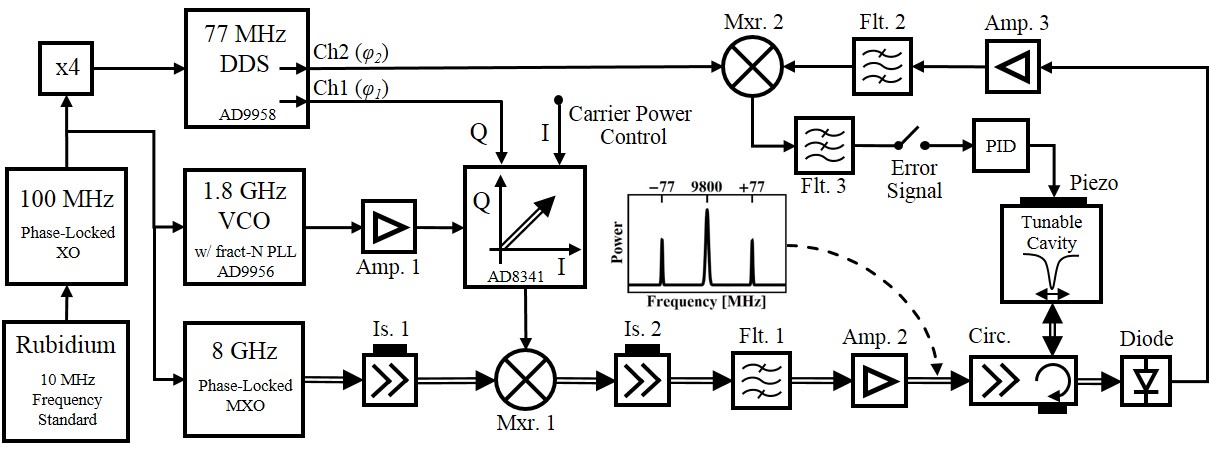}
\caption{\label{BlockDiagram}The cavity resonance frequency is measured with a microwave version of a Pound-Drever-Hall lock, which generates a signal proportional to the difference between the source and cavity frequencies.  This is used as an error signal in a feedback loop to lock the cavity to the source, transferring the stability of the rubidium frequency reference, and the precise, digital tunability of the microwave source, to the cavity length.  Double-lined arrows indicate the path of high frequency ($\geq 8$\GigaHertz) signals, while single-lined arrows indicate low frequency signals.  The inset indicates the spectrum of the generated signal: a carrier and two phase-modulation-like sidebands. Amp: Amplifier, Is: Isolator, Mxr: Mixer, Flt: Filter, Circ: Circulator.}
\end{figure*}

Figure \ref{LabeledAssembly} shows a cross section of the constructed mechanical assembly.  The adjustable length is realized by cutting the cavity along a plane perpendicular to the axis and equidistant from the ends.  There is no current flowing in the walls there and therefore the Q and the resonant frequency of the cavity are unaffected by a small gap.  The left half-cavity is tied to the end of the external tube housing the device - indicated as the ``Fixed'' Half in the figure - setting the coarse length of the cavity.  The holder of the ``Moving'' Half - attached to the right end of the housing tube - is variable length.  A stacked-piezo tube (Piezomechanic HPSt 150/14-10/12), kept under compressive load throughout its stroke, allows elastic elongation of the concentric aluminum tube over a 10\MicroMeter\ range.  The shape of the variable-length holder is designed to minimize strain on the cavity wall.  By applying a layer of damping material (not shown) around the flexure tube, the stability of the feedback loop could be extended from about $2$\KiloHertz\ to $8$\KiloHertz.  Both half-cavity pieces were fabricated out of aluminum 7075 and subsequently gold plated, resulting in a loaded Q of 1280.

While the 15\MilliMeter\ cavity length is set by the choice of the 10\GigaHertz\ working frequency, the 3.5\MilliMeter\ and 11\MilliMeter\ inner and outer diameter were chosen as a compromise between quality factor, coupling dynamic range, and other conveniences.  An oversized coaxial waveguide (whose inner diameter matches that of the cavity's inner conductor), ending with a half-toroidal curved wall, brings the microwaves close to the fixed half-cavity end wall, and a taper element allows connecting it to a standard SMA connector.  Coupling is done with an iris (small hole) on the cavity/waveguide wall, which is partially blocked by a floating copper pill held in place by a dielectric screw.  Moving the pill in front of the iris allows precise control of the coupling, over a range of ${0.8<\beta<1.2}$, where ${\beta=(1\pm|S|)/(1\mp|S|)}$, and $S$ is the scattering parameter on resonance; the top (bottom) sign is chosen for over- (under-) coupled cases.  We have found that a return loss greater than 50\DeciBell\ optimizes the signal-to-noise ratio in our system, necessitating a smoothly-working adjustable coupling.

To evaluate the motion generated by the piezo tube actuator, an optical interferometer was added along the axis of the cavity.  A single-mode-fiber-coupled 1310\NanoMeter\ laser diode was used as a light source.  Light was brought through a fiber-coupled circulator to a lens assembly consisting of a gradient index lens and a 3\MilliMeter\ radius plano-concave lens affixed to the moving half-cavity. The lens ensemble has a 3\MilliMeter\ working distance, assuring that light rays stepping to air from glass are perpendicular to the concave surface - which has no anti-reflection coating - providing a proper reference surface to a flat mirror placed at the beam waist. The light is focused onto the aluminum-coated end of a polished quartz rod held by the fixed half-cavity.  The lens collects the back-reflected light, as well as the ${\sim4\%}$ reflected off the concave glass-air interface and couples them back into the single-mode fiber.  The fiber-coupled circulator directs them into a photodiode; the observed interference signal reflects the distance change between the concave lens surface and the end of the quartz rod, allowing to monitor the relative motion of the half cavities.

The cavity resonant frequency is compared to that generated by a home-built microwave source, shown in the block diagram in figure \ref{BlockDiagram}.  We have implemented a microwave version of what is commonly known in optics as a Pound-Drever-Hall (PDH) lock\cite{SteinTurneaure1,*SteinTurneaure2,*MannBlair,*Drever2,*Black}.  Driving the piezo-actuator to match the cavity resonance frequency to that of the microwave signal, the cavity frequency inherits the stability and digital tunability of the source, stable to ${\sim10}$\MilliHertz, and adjustable in the range of ${9810\pm15}$\MegaHertz, with better than millihertz precision.

The frequency modulated 9.8\GigaHertz\ signal is generated in multiple steps.  All signal sources are phase-locked directly or indirectly to a rubidium frequency reference (SRS PRS10), ensuring $2 \times 10^{-12}$ stability over 100\Second.  The fine frequency control and modulation is implemented at 1.8\GigaHertz, and then converted up to 9.8\GigaHertz\ using an 8\GigaHertz\ low phase-noise source (Wenzel multiplied crystal oscillator (MXO)).  A ${1810\pm15}$\MegaHertz\ VCO (ZCommunications ZRO1820A1LF) is controlled via a fractional-N phase-lock-loop (PLL) that internally includes a direct digital synthesizer (DDS) chip with a 48-bit frequency tuning word (Analog Devices AD9956).  The reference for the PLL is a 100\MegaHertz\ crystal oscillator (Wenzel SC-cut XO) with low close-in phase noise.  This configuration provides digital frequency control with ${\sim11}$\MicroHertz\ resolution.  The VCO output is taken to a vector (\textbf{\textit{IQ}}) modulator (Analog Devices AD8341), which amplitude modulates (${\pm100\%}$, as in double-sideband suppressed-carrier transmission) the \textbf{\textit{Q}} component at a frequency of 77\MegaHertz.  The modulation frequency, generated by channel 1 of a two-channel DDS board (Analog Devices AD9958), is chosen to be much higher than the cavity half-width (3.8\MegaHertz) for reasons discussed below.  The \textbf{\textit{I}} component is not modulated, but optionally attenuated by setting the DC voltage on the \textbf{\textit{I}} baseband input.  The resulting output of the vector modulator is a carrier at 1.8\GigaHertz, and two sidebands at ${\pm77}$\MegaHertz.  This is converted up to 9.8\GigaHertz\ (figure \ref{BlockDiagram} inset), amplified to 12\dBm\ carrier and -12\dBm\ sideband power, and directed to the coarsely frequency-matched cavity through a circulator.  

The reflected carrier, being close to the cavity resonance frequency, is phase shifted proportionally with the frequency difference, undergoing a $\pi$ phase change over the cavity width.  The sidebands, chosen to be far from the cavity, are reflected with constant phase shift, and serve as a reference for measuring the phase-shift of the carrier.  After detection with a biased Schottky diode (Herotek DDS118), the beating of the sidebands with the carrier is selectively detected with the help of a mixer and the properly-phased reference, provided by the channel 2 output of the DDS board, to generate the error signal.  The error signal is fed-back to the cavity piezo through a PID amplifier to lock the cavity to the source.  An Asylum Research MFP3D controller conveniently provides several analog-to-digital and digital-to-analog converters, a -10 to +150\Volt\ high voltage amplifier, and digital PID loop needed to run and monitor the system.  Software to set power levels, frequencies, and/or phase of the various signal generators was written in Wavemetrics IGOR Pro.

The described system is used as an actuator in closed-loop active feedback mode, controlling the motion through setting the frequency value. The exponential settling time constant - limited by the mechanical motion - for a 5\KiloHertz\ (8\NanoMeter) step in the setpoint\footnote{This required using an analog PID (SRS SIM960) because of a 70\MicroSecond\ delay associated with the digital electronics.  For other measurements reported, the digital PID was used, limiting the bandwidth to 1\KiloHertz.  We checked that the settling time-constant of the VCO was faster than 20\MicroSecond.} is about 20\MicroSecond.  While adding payload, i.e. the mass of a sample holder, will reduce the bandwidth, we estimate that an X-Y scanner using this actuator would be able to collect up to 5 full-scale, high-resolution images per second.

\begin{figure}[tb]
\includegraphics[width=\linewidth]{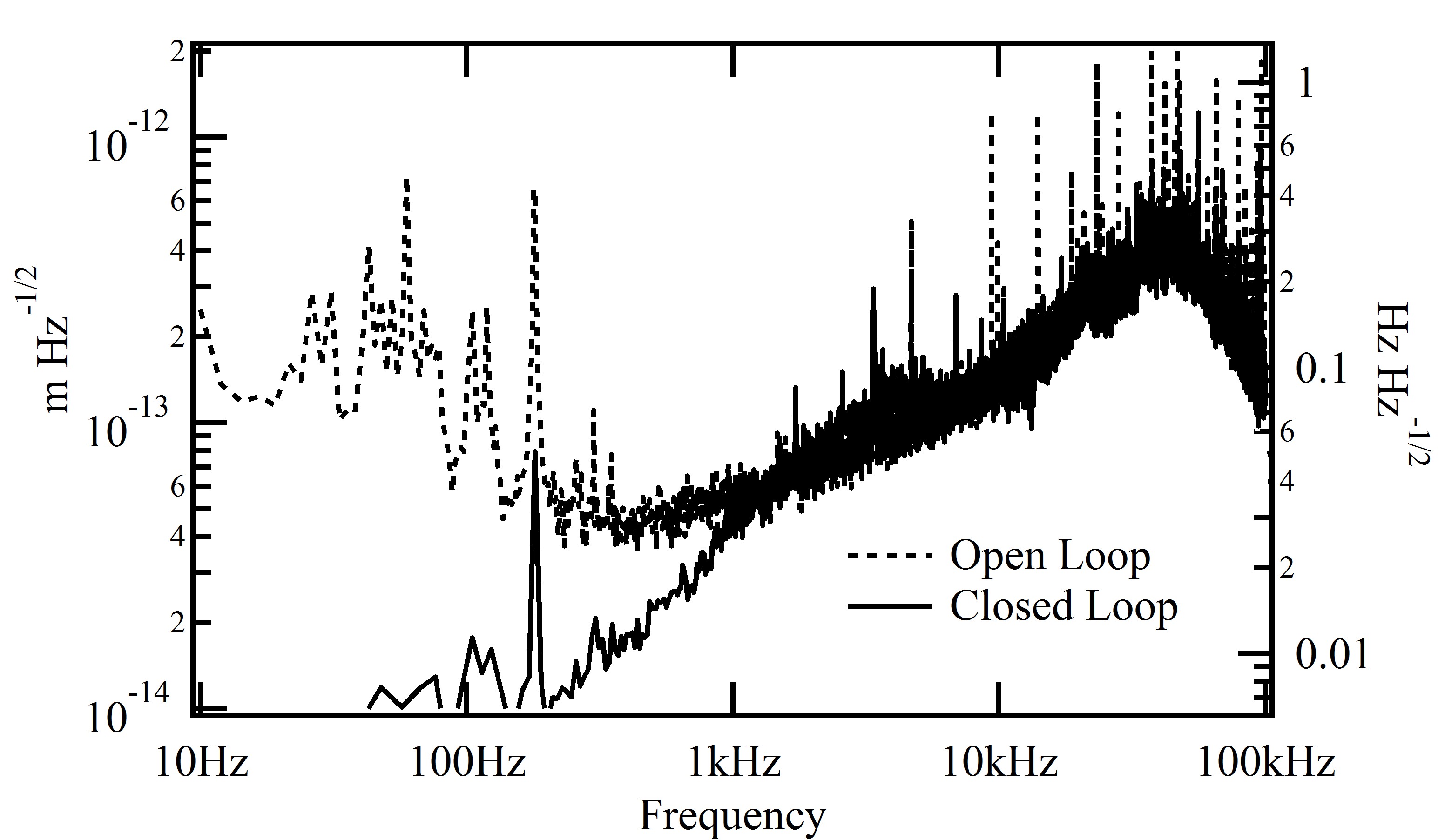}
\caption{\label{Noise} The noise spectral density of the open and closed loop PDH error signal.  Closing the loop stabilizes the cavity, and suppresses drift and vibration within the loop bandwidth.}
\end{figure}

The precision of the system is determined by the quality of the microwave sources used.  The frequency analysis of the error (PDH) signal is shown both for closed- and open-loop operation in figure \ref{Noise}.  The open-loop noise spectral density curve shows a minimum of $60$\FemtoMeterPerRtHz\ around 500\Hertz\ as the low frequency noise is dominated by the cavity frequency noise; we have made no particular effort to isolate vibrational and acoustical noise from the cavity.  As the mechanical noise diminishes at higher frequencies, the frequency difference noise starts to get dominated by the phase noise of the microwave signal sources.  Turning on the feedback loop actively compensates the vibration and acoustic noise, but also converts phase noise to length fluctuations within the loop bandwidth.  Above the loop bandwidth (1\KiloHertz\ in the figure), the closed loop signal practically matches the open loop signal.  From a noise and stability point-of-view, the optimal bandwidth is where the diminishing (in frequency) mechanical noise intersects the increasing frequency noise of the source.  In practice however, the bandwidth is chosen as a compromise between RMS noise and settling time.  Limiting the feedback bandwidth to 1\KiloHertz, and taking the distance noise to be 60\FemtoMeterPerRtHz\ everywhere in this range allows us to claim a less than 2\PicoMeter\ RMS noise in a 1\KiloHertz\ bandwidth.

\begin{figure}[tb]
\includegraphics[width=\linewidth]{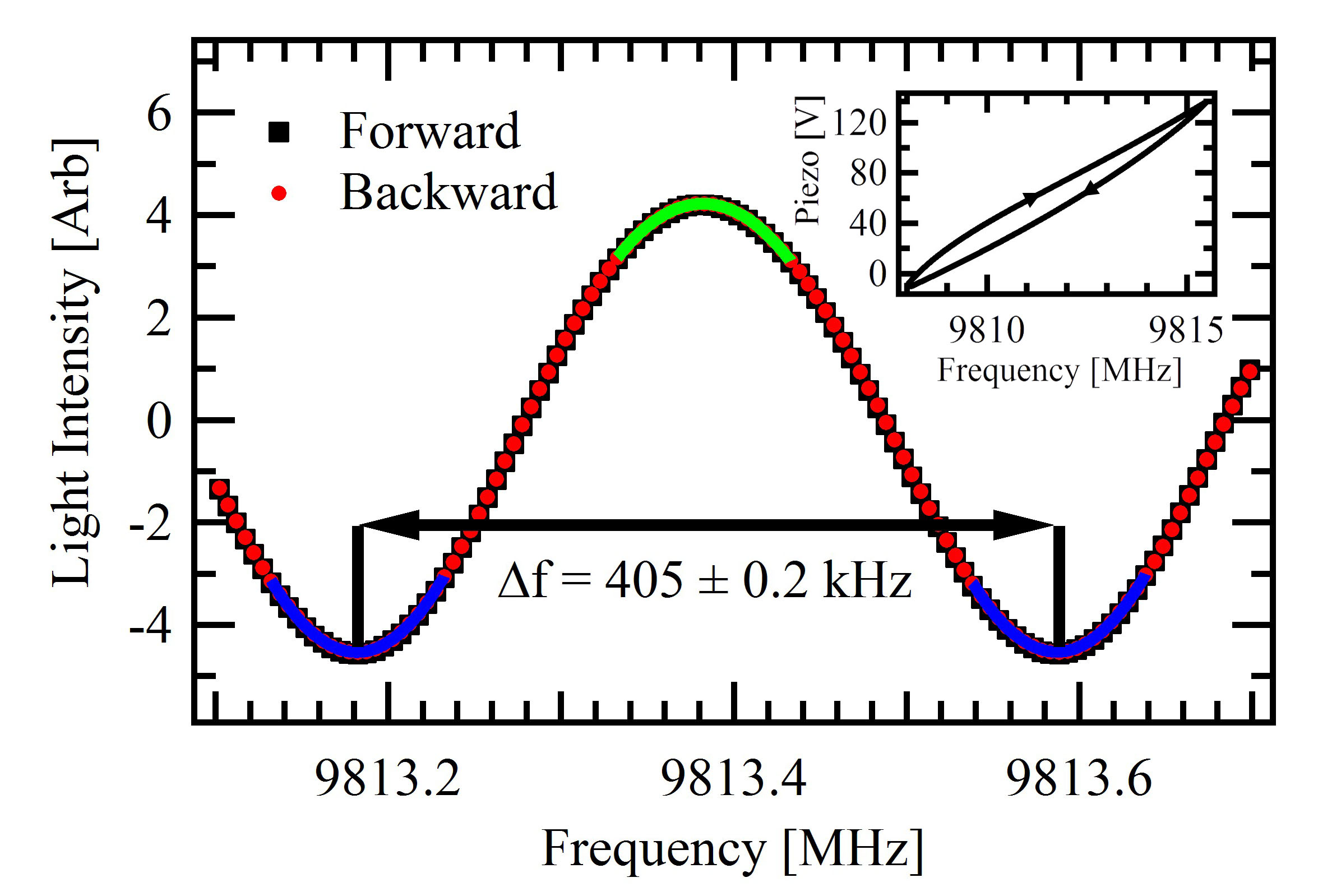}
\caption{\label{fig3}(Color online) The $\lambda /2$ periodicity of the interferometer signal as the cavity frequency is scanned in closed-loop is used to verify that the relationship  $\Delta L=-c \Delta f / 2f^2$ is obeyed in a limited range.  We verified a 5\% deviation at a 120\MicroMeter\ gap between the cavity halves. The hysteresis of the piezo is displayed in the inset.}
\end{figure}

Obviously, the fiber interferometer built into the middle of the cavity was never meant to match the precision claimed, but it was useful to characterize the overall system behavior, such as trouble shooting, verifying scan range, piezo-hysteresis, etc.  Figure \ref{fig3} illustrates the limitations of the interferometer measurement, showing subsequent fringes obtained from a full range scan.  We fit parabolas to locate successive minima or maxima (blue and green curves) and extracted the frequency difference corresponding to the $\lambda/2$ periodicity with $\pm0.5$\% precision, i.e. about $\pm 3$\NanoMeter, reproducibly. We averaged the fringe spacing from an equal number of forward and backward scans (10 each) to compensate for drifts.  

The ${\pm0.5\%}$ irreproducibility is due to small changes in the alignment and coarse length/frequency adjustment upon rebuilding the assembly.  This precision would not allow us to detect any deviations from the ideal relation $f=c/2L$ if the gap between the cavity halves is less than 10\MicroMeter.  Our differential measurement would be indistinguishable from $\Delta L=-c \Delta f / 2f^2$.  To verify the useful range of our device, we have simulated the cavity frequency as a function of the gap width in Ansys HFSS.  The quadratic correction results in a 1\% and 5\% deviation in the linearized frequency-distance relationship at 45 and 120\MicroMeter\ respectively.  The 5\% deviation with a 120\MicroMeter\ gap was sufficiently significant to verify with our interferometer within the experimental error  (figure \ref{fig3}).  The back extrapolation of the values confirms that less than 0.1\% deviation from the ideal slope is expected in the 0-10\MicroMeter\ range, hence, the integrated error in the absolute distance over the 10\MicroMeter\ full range would be less than $10^{-4}$.  Consequently atomic scale distances as well as periodicities (revealed by Fourier transformation of data collected over the full range) would be determined with better than 1\% precision, comparable to that of lattice parameters obtained by bulk crystallography.

Incorporating such an encoder into the positioning stage of an SPM would require two encoders to operate in perpendicular directions with minimal cross-talk.  To capitalize on the stability offered by frequency standards, stringent environmental control is needed to eliminate temperature variation of the cavity wall resistivity\footnote{Drift of about 2.5\NmPerK\ can be expected based on the thermal coefficient of resistivity for common, high-conductivity metals.}, the dielectric constant of the gas within\footnote{It was observed that thermal and humidity fluctuations caused the largest changes in the dielectric constant of the air within the cavity sitting in atmosphere\cite{Thesis}.}, just as well as the rest of the scanner to limit thermal drift of all mechanical parts involved.  The effective ``thermal coefficient'' of the encoder is expected to be $1.7\times10^{-7}$\PerK\ (for 15\MilliMeter\ cavity length), almost an order of magnitude smaller than that of Invar.

We note that the precision of any other Automatic Frequency Control (AFC) method\cite{PoundLock} would also be ultimately limited by the phase stability of the source.  The realized PDH-like system advantage is a fast response time allowing to explore the limitations of our mechanical design.

We have demonstrated that compact, coaxial microwave cavities make excellent displacement measurement tools, capable of picometer precision.  The described device has a range of 10\MicroMeter, limited by the chosen piezo-actuator, and a noise floor of 60\FemtoMeterPerRtHz, limited by the quality of the signal generators.  The actuator system is capable of self-calibration to an accuracy of better than 0.1\% of the displacement range, although for ranges larger than 10s of micrometer, a non-linear correction will be needed to account for the larger gap.  Incorporating displacement encoders using these principles into SPM would go a considerable way towards realization of a metrological SPM.

\begin{acknowledgments}
We thank Emil Kirilov for valuable discussions, Asylum Research for electronics and technical support, and the UCLA physics machine shop for fabricating the intricate pieces in this work.\\
\end{acknowledgments}

\bibliography{DisplacementEncoder}

\begin{thebibliography}{15}%
\makeatletter
\providecommand \@ifxundefined [1]{%
 \@ifx{#1\undefined}
}%
\providecommand \@ifnum [1]{%
 \ifnum #1\expandafter \@firstoftwo
 \else \expandafter \@secondoftwo
 \fi
}%
\providecommand \@ifx [1]{%
 \ifx #1\expandafter \@firstoftwo
 \else \expandafter \@secondoftwo
 \fi
}%
\providecommand \natexlab [1]{#1}%
\providecommand \enquote  [1]{``#1''}%
\providecommand \bibnamefont  [1]{#1}%
\providecommand \bibfnamefont [1]{#1}%
\providecommand \citenamefont [1]{#1}%
\providecommand \href@noop [0]{\@secondoftwo}%
\providecommand \href [0]{\begingroup \@sanitize@url \@href}%
\providecommand \@href[1]{\@@startlink{#1}\@@href}%
\providecommand \@@href[1]{\endgroup#1\@@endlink}%
\providecommand \@sanitize@url [0]{\catcode `\\12\catcode `\$12\catcode
  `\&12\catcode `\#12\catcode `\^12\catcode `\_12\catcode `\%12\relax}%
\providecommand \@@startlink[1]{}%
\providecommand \@@endlink[0]{}%
\providecommand \url  [0]{\begingroup\@sanitize@url \@url }%
\providecommand \@url [1]{\endgroup\@href {#1}{\urlprefix }}%
\providecommand \urlprefix  [0]{URL }%
\providecommand \Eprint [0]{\href }%
\providecommand \doibase [0]{http://dx.doi.org/}%
\providecommand \selectlanguage [0]{\@gobble}%
\providecommand \bibinfo  [0]{\@secondoftwo}%
\providecommand \bibfield  [0]{\@secondoftwo}%
\providecommand \translation [1]{[#1]}%
\providecommand \BibitemOpen [0]{}%
\providecommand \bibitemStop [0]{}%
\providecommand \bibitemNoStop [0]{.\EOS\space}%
\providecommand \EOS [0]{\spacefactor3000\relax}%
\providecommand \BibitemShut  [1]{\csname bibitem#1\endcsname}%
\let\auto@bib@innerbib\@empty
\bibitem [{\citenamefont {Mann}\ and\ \citenamefont {Blair}(1983)}]{MannBlair}%
  \BibitemOpen
  \bibfield  {author} {\bibinfo {author} {\bibfnamefont {A.~G.}\ \bibnamefont
  {Mann}}\ and\ \bibinfo {author} {\bibfnamefont {D.~G.}\ \bibnamefont
  {Blair}},\ }\href@noop {} {\bibfield  {journal} {\bibinfo  {journal} {Journal
  of Physics D: Applied Physics}\ }\textbf {\bibinfo {volume} {16}},\ \bibinfo
  {pages} {105} (\bibinfo {year} {1983})}\BibitemShut {NoStop}%
\bibitem [{\citenamefont {Blair}, \citenamefont {Ivanov},\ and\ \citenamefont
  {Peng}(1992)}]{BlairIvanovPeng}%
  \BibitemOpen
  \bibfield  {author} {\bibinfo {author} {\bibfnamefont {D.~G.}\ \bibnamefont
  {Blair}}, \bibinfo {author} {\bibfnamefont {E.~N.}\ \bibnamefont {Ivanov}}, \
  and\ \bibinfo {author} {\bibfnamefont {H.}~\bibnamefont {Peng}},\ }\href@noop
  {} {\bibfield  {journal} {\bibinfo  {journal} {Journal of Physics D: Applied
  Physics}\ }\textbf {\bibinfo {volume} {25}},\ \bibinfo {pages} {1110}
  (\bibinfo {year} {1992})}\BibitemShut {NoStop}%
\bibitem [{\citenamefont {Cuthbertson}\ \emph {et~al.}(1998)\citenamefont
  {Cuthbertson}, \citenamefont {Tobar}, \citenamefont {Ivanov},\ and\
  \citenamefont {Blair}}]{CuthbertsonTobarIvanovBlair}%
  \BibitemOpen
  \bibfield  {author} {\bibinfo {author} {\bibfnamefont {B.~D.}\ \bibnamefont
  {Cuthbertson}}, \bibinfo {author} {\bibfnamefont {M.~E.}\ \bibnamefont
  {Tobar}}, \bibinfo {author} {\bibfnamefont {E.~N.}\ \bibnamefont {Ivanov}}, \
  and\ \bibinfo {author} {\bibfnamefont {D.~G.}\ \bibnamefont {Blair}},\
  }\href@noop {} {\bibfield  {journal} {\bibinfo  {journal} {IEEE Transactions
  on Ultrasonics, Ferroelectrics, and Frequency Control}\ }\textbf {\bibinfo
  {volume} {45}},\ \bibinfo {pages} {1303} (\bibinfo {year}
  {1998})}\BibitemShut {NoStop}%
\bibitem [{\citenamefont {Zhu}\ \emph {et~al.}(2006)\citenamefont {Zhu},
  \citenamefont {McElroy}, \citenamefont {Lee}, \citenamefont {Devereaux},
  \citenamefont {Si}, \citenamefont {Davis},\ and\ \citenamefont
  {Balatsky}}]{PhysRevLett.97.177001}%
  \BibitemOpen
  \bibfield  {author} {\bibinfo {author} {\bibfnamefont {J.-X.}\ \bibnamefont
  {Zhu}}, \bibinfo {author} {\bibfnamefont {K.}~\bibnamefont {McElroy}},
  \bibinfo {author} {\bibfnamefont {J.}~\bibnamefont {Lee}}, \bibinfo {author}
  {\bibfnamefont {T.~P.}\ \bibnamefont {Devereaux}}, \bibinfo {author}
  {\bibfnamefont {Q.}~\bibnamefont {Si}}, \bibinfo {author} {\bibfnamefont
  {J.~C.}\ \bibnamefont {Davis}}, \ and\ \bibinfo {author} {\bibfnamefont
  {A.~V.}\ \bibnamefont {Balatsky}},\ }\href {\doibase
  10.1103/PhysRevLett.97.177001} {\bibfield  {journal} {\bibinfo  {journal}
  {Phys. Rev. Lett.}\ }\textbf {\bibinfo {volume} {97}},\ \bibinfo {pages}
  {177001} (\bibinfo {year} {2006})}\BibitemShut {NoStop}%
\bibitem [{\citenamefont {Hanaguri}\ \emph {et~al.}(2007)\citenamefont
  {Hanaguri}, \citenamefont {Kohsaka}, \citenamefont {Davis}, \citenamefont
  {Lupien}, \citenamefont {Yamada}, \citenamefont {Azuma}, \citenamefont
  {Takano}, \citenamefont {Ohishi}, \citenamefont {Ono},\ and\ \citenamefont
  {Takagi}}]{ISI:000251456900024}%
  \BibitemOpen
  \bibfield  {author} {\bibinfo {author} {\bibfnamefont {T.}~\bibnamefont
  {Hanaguri}}, \bibinfo {author} {\bibfnamefont {Y.}~\bibnamefont {Kohsaka}},
  \bibinfo {author} {\bibfnamefont {J.~C.}\ \bibnamefont {Davis}}, \bibinfo
  {author} {\bibfnamefont {C.}~\bibnamefont {Lupien}}, \bibinfo {author}
  {\bibfnamefont {I.}~\bibnamefont {Yamada}}, \bibinfo {author} {\bibfnamefont
  {M.}~\bibnamefont {Azuma}}, \bibinfo {author} {\bibfnamefont
  {M.}~\bibnamefont {Takano}}, \bibinfo {author} {\bibfnamefont
  {K.}~\bibnamefont {Ohishi}}, \bibinfo {author} {\bibfnamefont
  {M.}~\bibnamefont {Ono}}, \ and\ \bibinfo {author} {\bibfnamefont
  {H.}~\bibnamefont {Takagi}},\ }\href@noop {} {\bibfield  {journal} {\bibinfo
  {journal} {NATURE PHYSICS}\ }\textbf {\bibinfo {volume} {3}},\ \bibinfo
  {pages} {865} (\bibinfo {year} {2007})}\BibitemShut {NoStop}%
\bibitem [{\citenamefont {Pisani}\ \emph {et~al.}(2012)\citenamefont {Pisani},
  \citenamefont {Yacoot}, \citenamefont {Balling}, \citenamefont {Bancone},
  \citenamefont {Birlikseven}, \citenamefont {Celik}, \citenamefont {Fluegge},
  \citenamefont {Hamid}, \citenamefont {Koechert}, \citenamefont {Kren},
  \citenamefont {Kuetgens}, \citenamefont {Lassila}, \citenamefont {Picotto},
  \citenamefont {Sahin}, \citenamefont {Seppa}, \citenamefont {Tedaldi},\ and\
  \citenamefont {Weichert}}]{ISI:000306831400006}%
  \BibitemOpen
  \bibfield  {author} {\bibinfo {author} {\bibfnamefont {M.}~\bibnamefont
  {Pisani}}, \bibinfo {author} {\bibfnamefont {A.}~\bibnamefont {Yacoot}},
  \bibinfo {author} {\bibfnamefont {P.}~\bibnamefont {Balling}}, \bibinfo
  {author} {\bibfnamefont {N.}~\bibnamefont {Bancone}}, \bibinfo {author}
  {\bibfnamefont {C.}~\bibnamefont {Birlikseven}}, \bibinfo {author}
  {\bibfnamefont {M.}~\bibnamefont {Celik}}, \bibinfo {author} {\bibfnamefont
  {J.}~\bibnamefont {Fluegge}}, \bibinfo {author} {\bibfnamefont
  {R.}~\bibnamefont {Hamid}}, \bibinfo {author} {\bibfnamefont
  {P.}~\bibnamefont {Koechert}}, \bibinfo {author} {\bibfnamefont
  {P.}~\bibnamefont {Kren}}, \bibinfo {author} {\bibfnamefont {U.}~\bibnamefont
  {Kuetgens}}, \bibinfo {author} {\bibfnamefont {A.}~\bibnamefont {Lassila}},
  \bibinfo {author} {\bibfnamefont {G.~B.}\ \bibnamefont {Picotto}}, \bibinfo
  {author} {\bibfnamefont {E.}~\bibnamefont {Sahin}}, \bibinfo {author}
  {\bibfnamefont {J.}~\bibnamefont {Seppa}}, \bibinfo {author} {\bibfnamefont
  {M.}~\bibnamefont {Tedaldi}}, \ and\ \bibinfo {author} {\bibfnamefont
  {C.}~\bibnamefont {Weichert}},\ }\href@noop {} {\bibfield  {journal}
  {\bibinfo  {journal} {Metrologia}\ }\textbf {\bibinfo {volume} {49}},\
  \bibinfo {pages} {455} (\bibinfo {year} {2012})}\BibitemShut {NoStop}%
\bibitem [{\citenamefont {Stein}\ and\ \citenamefont
  {Turneaure}(1972)}]{SteinTurneaure1}%
  \BibitemOpen
  \bibfield  {author} {\bibinfo {author} {\bibfnamefont {S.~R.}\ \bibnamefont
  {Stein}}\ and\ \bibinfo {author} {\bibfnamefont {J.~P.}\ \bibnamefont
  {Turneaure}},\ }\href@noop {} {\bibfield  {journal} {\bibinfo  {journal}
  {Electronics Letters}\ }\textbf {\bibinfo {volume} {8}},\ \bibinfo {pages}
  {321} (\bibinfo {year} {1972})}\BibitemShut {NoStop}%
\bibitem [{\citenamefont {Stein}\ and\ \citenamefont
  {Turneaure}(1973)}]{SteinTurneaure2}%
  \BibitemOpen
  \bibfield  {author} {\bibinfo {author} {\bibfnamefont {S.~R.}\ \bibnamefont
  {Stein}}\ and\ \bibinfo {author} {\bibfnamefont {J.~P.}\ \bibnamefont
  {Turneaure}},\ }in\ \href@noop {} {\emph {\bibinfo {booktitle} {27th Annual
  Frequency Control Symposium}}}\ (\bibinfo {year} {1973})\ pp.\ \bibinfo
  {pages} {414--420}\BibitemShut {NoStop}%
\bibitem [{\citenamefont {Drever}\ \emph {et~al.}(1983)\citenamefont {Drever},
  \citenamefont {Hall}, \citenamefont {Kowalski}, \citenamefont {Hough},
  \citenamefont {Ford}, \citenamefont {Munley},\ and\ \citenamefont
  {Ward}}]{Drever2}%
  \BibitemOpen
  \bibfield  {author} {\bibinfo {author} {\bibfnamefont {R.~W.~P.}\
  \bibnamefont {Drever}}, \bibinfo {author} {\bibfnamefont {J.~L.}\
  \bibnamefont {Hall}}, \bibinfo {author} {\bibfnamefont {F.~V.}\ \bibnamefont
  {Kowalski}}, \bibinfo {author} {\bibfnamefont {J.}~\bibnamefont {Hough}},
  \bibinfo {author} {\bibfnamefont {G.~M.}\ \bibnamefont {Ford}}, \bibinfo
  {author} {\bibfnamefont {A.~J.}\ \bibnamefont {Munley}}, \ and\ \bibinfo
  {author} {\bibfnamefont {H.}~\bibnamefont {Ward}},\ }\href@noop {} {\bibfield
   {journal} {\bibinfo  {journal} {Applied Physics B}\ }\textbf {\bibinfo
  {volume} {31}},\ \bibinfo {pages} {97} (\bibinfo {year} {1983})}\BibitemShut
  {NoStop}%
\bibitem [{\citenamefont {Black}(2001)}]{Black}%
  \BibitemOpen
  \bibfield  {author} {\bibinfo {author} {\bibfnamefont {E.~D.}\ \bibnamefont
  {Black}},\ }\href@noop {} {\bibfield  {journal} {\bibinfo  {journal}
  {American Journal of Physics}\ }\textbf {\bibinfo {volume} {69}},\ \bibinfo
  {pages} {79} (\bibinfo {year} {2001})}\BibitemShut {NoStop}%
\bibitem [{Note1()}]{Note1}%
  \BibitemOpen
  \bibinfo {note} {This required using an analog PID (SRS SIM960) because of a
  70\penalty \@M \unhbox \voidb@x \hbox {$\protect \tmspace +\thickmuskip
  {.2777em}$\textmu s}\ delay associated with the digital electronics. For
  other measurements reported, the digital PID was used, limiting the bandwidth
  to 1\penalty \@M \unhbox \voidb@x \hbox {$\protect \tmspace +\thickmuskip
  {.2777em}$kHz}. We checked that the settling time-constant of the VCO was
  faster than 20\penalty \@M \unhbox \voidb@x \hbox {$\protect \tmspace
  +\thickmuskip {.2777em}$\textmu s}.}\BibitemShut {Stop}%
\bibitem [{Note2()}]{Note2}%
  \BibitemOpen
  \bibinfo {note} {Drift of about 2.5\penalty \@M \unhbox \voidb@x \hbox
  {$\protect \tmspace +\thickmuskip {.2777em}$nm$\protect \tmspace
  +\thickmuskip {.2777em}$K$^{-1}$}\ can be expected based on the thermal
  coefficient of resistivity for common, high-conductivity metals.}\BibitemShut
  {Stop}%
\bibitem [{Note3()}]{Note3}%
  \BibitemOpen
  \bibinfo {note} {It was observed that thermal and humidity fluctuations
  caused the largest changes in the dielectric constant of the air within the
  cavity sitting in atmosphere\cite {Thesis}.}\BibitemShut {Stop}%
\bibitem [{\citenamefont {Pound}(1946)}]{PoundLock}%
  \BibitemOpen
  \bibfield  {author} {\bibinfo {author} {\bibfnamefont {R.~V.}\ \bibnamefont
  {Pound}},\ }\href@noop {} {\bibfield  {journal} {\bibinfo  {journal} {The
  Review of Scientific Instruments}\ }\textbf {\bibinfo {volume} {17}},\
  \bibinfo {pages} {490} (\bibinfo {year} {1946})}\BibitemShut {NoStop}%
\bibitem [{\citenamefont {Koulakis}(2014)}]{Thesis}%
  \BibitemOpen
  \bibfield  {author} {\bibinfo {author} {\bibfnamefont {J.~P.}\ \bibnamefont
  {Koulakis}},\ }\href
  {http://search.proquest.com/docview/1524690403?accountid=14512}
  {{\enquote {\bibinfo {title} {Traceable and precise
  displacement measurements with microwave cavities},}\ }} (\bibinfo {year}
  {2014}),\ \bibinfo {note} {copyright ProQuest, UMI Dissertations Publishing
  2014;}\BibitemShut {NoStop}%
\end{thebibliography}%
\end{document}